\documentclass[a4paper,10pt,twoside]{cpc-hepnp}
\usepackage{CJK,upgreek,fancyhdr}
\usepackage{multicol}
\usepackage{graphicx}
\usepackage{booktabs}
\usepackage{amssymb,bm,mathrsfs,bbm,amscd}
\usepackage[tbtags]{amsmath}
\usepackage{lastpage}

\begin{document}
\begin{CJK*}{GBK}{song}

\fancyhead[c]{\small Chinese Physics C~~~Vol. xx, No. x (201x) xxxxxx}
\fancyfoot[C]{\small xxxxxx-\thepage}

\footnotetext[0]{Received 15 November 2018}

\title{Charged scalar fields in a Kerr-Sen black hole: exact solutions, Hawking radiation, and resonant frequencies\thanks{H.S.V. is funded by the CNPq through the research Project No. 150640/2018-8. V.B.B. is partially supported by the CNPq through the research Project No. 305835/2016-5}}

\author{%
      H. S. Vieira$^{1}$\email{horacio.santana.vieira@hotmail.com}%
\quad	V. B. Bezerra$^{1}$\email{valdir@fisica.ufpb.br}
}
\maketitle

\address{%
$^{1}$ Departamento de F\'{i}sica, Universidade Federal da Para\'{i}ba, Caixa Postal 5008, CEP 58051-970, Jo\~{a}o Pessoa, PB, Brazil
}

\begin{abstract}
In this study, we consider charged massive scalar fields around a Kerr-Sen spacetime. The radial and angular parts of the covariant Klein-Gordon equation are solved in terms of the confluent Heun function. From the exact radial solution, we obtain the Hawking radiation spectrum and discuss its resonant frequencies. The massless case of the resonant frequencies is also examined.
\end{abstract}



\footnotetext[0]{\hspace*{-3mm}\raisebox{0.3ex}{$\scriptstyle\copyright$}2013
Chinese Physical Society and the Institute of High Energy Physics
of the Chinese Academy of Sciences and the Institute
of Modern Physics of the Chinese Academy of Sciences and IOP Publishing Ltd}%

%
%
\section{Introduction}
Black holes are interesting objects predicted by the theory of general relativity. To understand the physics of these objects, we can study the physical processes that occur in spacetime associated with them. Among these studies, we can mention the ones corresponding to Hawking radiation, scattering of particles and fields of different spins, and resonant frequencies \cite{LettNuovoCimento.15.257,PhysRevD.16.937,PhysRevD.18.1030,PhysRevD.22.2323,ZhEkspTeorFiz.92.369,PhysRevD.75.104012,IntJModPhysD.21.1250045,CommunTheorPhys.62.227,PhysRevD.96.024011}. The last one is of special interest to us and will be examined in this study on scalar fields.

The investigation of these phenomena involves the solution of an ordinary differential equation. In our case, we want to solve the Klein-Gordon equation for a charged massive scalar field in the background under consideration. Thus, to obtain the exact solution for this equation, we need to use the most general function in mathematical physics, namely, Heun functions. During last decades, Heun functions have gained increasingly more importance due to its large number of applications in different areas of physics and mathematics. Otherwise, without the use of these functions, finding the exact analytical solutions of the Klein-Gordon equation in the entire spacetime is not possible \cite{ClassQuantumGrav.31.045003,arXiv:1101.0471v8,AnnPhys.395.138,AdHighEnergyPhys.2018.8621573,AppMathComp.338.624}.

Recently, several physical processes in the background of Kerr-Sen black holes were studied, for instance, the scattering of photons and the evaporation process. Furthermore, the conformal symmetries of a Kerr-Sen black hole and the instability of a bound state for a charged massive scalar field in the spacetime background of such a black hole is studied. In addition, the holographic dual of a conformal field theory (CFT) for the scattering process in the background of a Kerr-Sen black hole is investigated. Other important issues regarding the Kerr-Sen background are as follows: properties of shadow, cosmic censorship conjecture, Hawking-temperature and entropy, and the increasing amount of evidence that various galaxies contain a supermassive black hole at their center. Therefore, these lines of research have motivated researchers to develop theoretical approaches in order to explain such phenomena \cite{arXiv:1610.09477,PhysRevD.96.105025,ClassQuantumGrav.34.155002,PhysRevD.95.124050,ModPhysLett.A33.1850099,PhysLettB.782.594,PhysLettB.782.185,PhysRevD.97.024003,ClassQuantumGrav.35.025003}.

The thermal radiation of black holes describes black body radiation as being a quantum effect, as predicted in the mid-1970s \cite{Nature.248.30,CommunMathPhys.43.199}. It constitutes one of the most important characteristic which arises as an effect of the curvature of the spacetime on quantum fields. Thus, it is essentially a semiclassical effect due ti the fact that it occurs in a spacetime generated by classical gravitational fields which the matter fields are quantized. This implies that the Hawking radiation connects classical gravity and quantum field theory and hence its investigation could help us to understand gravity itself in addition to a better understanding of the physics of a black hole \cite{PhysRevD.13.2188,PhysRevD.14.332,PhysRevLett.85.5042,PhysRevLett.95.011303}. Therefore, it is important, from different perspectives, to investigate the Hawking radiation emitted by black holes.

While studying the physics of black holes, we can obtain information through the interaction of these fields with different quantum fields. One of these information are related to the resonant frequencies of the radiation emitted by black holes associated with exponentially decaying oscillations whose values depend on the parameters describing the black holes, such as mass, charge and angular momentum. Thus, calculating and analyzing these frequencies to obtain some information about the physics of these objects is important.

In this paper, we use the confluent Heun functions to obtain the exact solution of the Klein-Gordon equation for a charged massive scalar field in the Kerr-Sen spacetime. This solution is used to study the resonant frequency and Hawking radiation. While discussing the Hawking radiation, we study some aspects of the thermodynamics of this black hole.

The outline of this paper is as follows. In Section 2, we present the solutions of the Klein-Gordon equation for a charged massive scalar field in the Kerr-Sen black hole spacetime for both angular and radial parts. In Section 3, we discuss the Hawking radiation of scalar waves. In Section 4, we obtain the resonant frequencies for both massive and massless scalar particles. Finally, in Section 5, the conclusions are provided.
%
%
\section{Scalar fields in the Kerr-Sen black hole}\label{Sec.II}
We want to study the interaction between scalar fields and the background corresponding to the four-dimensional charged and rotating black hole solution in the low energy limit of the heterotic string field theory, called Kerr-Sen spacetime (KS spacetime), whose line element in the Boyer-Lindquist coordinates is given by \cite{PhysRevLett.69.1006}
\begin{eqnarray}
ds^{2} & = & -\frac{1}{\rho_{\mbox{\tiny{KS}}}^{2}}(\Delta_{\mbox{\tiny{KS}}}-a^{2}\sin^{2}\theta)\ dt^{2}+\frac{\rho_{\mbox{\tiny{KS}}}^{2}}{\Delta_{\mbox{\tiny{KS}}}}\ dr^{2}+\rho_{\mbox{\tiny{KS}}}^{2}\ d\theta^{2}\nonumber\\
&& +\biggl(\rho_{\mbox{\tiny{KS}}}^{2}+a^{2}\sin^{2}\theta+\frac{2Mra^{2}\sin^{2}\theta}{\rho_{\mbox{\tiny{KS}}}^{2}}\biggr)\sin^{2}\theta\ d\phi^{2}\nonumber\\
&& -\frac{4Mra}{\rho_{\mbox{\tiny{KS}}}^{2}}\sin^{2}\theta\ dt\ d\phi\ ,
\label{eq:metrica_KS}
\end{eqnarray}
with
\begin{equation}
\Delta_{\mbox{\tiny{KS}}}=r^{2}+2(b-M)r+a^{2}=(r-r_{+})(r-r_{-})\ ,
\label{eq:Delta_KS}
\end{equation}
\begin{equation}
r_{\pm}=(M-b)\pm\sqrt{(M-b)^{2}-a^{2}}\ ,
\label{eq:horizon_KS}
\end{equation}
\begin{equation}
\rho_{\mbox{\tiny{KS}}}^{2}=r^{2}+2br+a^{2}\cos^{2}\theta\ ,
\label{eq:rho_KS}
\end{equation}
\begin{equation}
b=\frac{Q^{2}}{2M}\ ,
\label{eq:b_KS}
\end{equation}
where $M$, $Q$, and $a=J/M$ denote the mass, charge, and angular momentum per unit mass of the KS black hole, respectively. The parameter $b$ is related to the dilaton field.

The four-vector electromagnetic potential is given by \cite{PhysRevD.94.085007}
\begin{equation}
A_{\sigma}=\biggl(-\frac{Qr}{\rho_{\mbox{\tiny{KS}}}^{2}},0,0,\frac{Qra\sin^{2}\theta}{\rho_{\mbox{\tiny{KS}}}^{2}}\biggr)\ .
\label{eq:potencial_EM_KS}
\end{equation}

In what follows, let us obtain some geometrical and thermodynamics parameters of the KS spacetime, which will be used in the next section.

Firstly, let us calculate the gravitational acceleration on the KS black hole horizon surface, $\kappa_{\pm}$, which is given by
\begin{equation}
\kappa_{\pm}=\frac{1}{2(r_{\pm}^{2}+2br_{\pm}+a^{2})}\left.\frac{d\Delta_{\mbox{\tiny{KS}}}}{dr}\right|_{r=r_{\pm}}=\frac{r_{\pm}-r_{\mp}}{2(r_{\pm}^{2}+2br_{\pm}+a^{2})}\ .
\label{eq:Gravitational_acceleration_KS}
\end{equation}

As for the Hawking temperature, $T_{\pm}$, we have
\begin{equation}
T_{\pm}=\frac{\kappa_{\pm}\hbar}{2\pi k_{B}}\ ,
\label{eq:Hawking_temperature_KS}
\end{equation}
where the thermodynamic quantity $\beta_{\pm}$ is given by
\begin{equation}
\beta_{\pm}=\frac{1}{k_{B}T_{\pm}}\ .
\label{eq:beta_KS}
\end{equation}

The surface areas of the exterior and interior event horizons, $\mathcal{A}_{\pm}$, are given by
\begin{equation}
\mathcal{A}_{\pm}=\left.\int\int\sqrt{-g}\ d\theta\ d\phi \right|_{r=r_{\pm}}=4 \pi r_{\pm}(r_{\pm}-2b)\ ,
\label{eq:area_KS}
\end{equation}
so that the entropy at the event horizon, $S_{\pm}$, can be written as
\begin{equation}
S_{\pm}=\frac{\mathcal{A}_{\pm}}{4}=\pi r_{\pm}(r_{\pm}-2b)\ .
\label{eq:entropia_KS}
\end{equation}

The dragging angular velocity, $\Omega_{\pm}$, and the electric potential, $\Phi_{\pm}$, are very near the event horizon $r_{\pm}$, and are given by
\begin{equation}
\Omega_{\pm}=-\left.\frac{g_{t\phi}}{g_{\phi\phi}}\right|_{r=r_{\pm}}=\frac{a}{r_{\pm}^{2}+2br_{\pm}+a^{2}}\ ,
\label{eq:angular_velocity_KS}
\end{equation}
\begin{equation}
\Phi_{\pm}=\frac{Qr_{\pm}}{r_{\pm}^{2}+2br_{\pm}+a^{2}}\ .
\label{eq:pot_ele_KS}
\end{equation}

Note that all these quantities depend on the event horizons that are determined by the parameters characterizing the black hole and the dilaton.
%
%
\subsection{Klein-Gordon equation}
Now, to study the behavior of a charged massive scalar field in a curved spacetime, we need to consider the covariant Klein-Gordon equation, which can be written as
\begin{eqnarray}
&& \biggl[\frac{1}{\sqrt{-g}}\partial_{\sigma}(g^{\sigma\tau}\sqrt{-g}\partial_{\tau})-ie(\partial_{\sigma}A^{\sigma})-2ieA^{\sigma}\partial_{\sigma}\nonumber\\
&& -\frac{ie}{\sqrt{-g}}A^{\sigma}(\partial_{\sigma}\sqrt{-g})-e^{2}A^{\sigma}A_{\sigma}-\mu_{0}^{2}\biggr]\Psi=0\ ,
\label{eq:Klein-Gordon_gauge}
\end{eqnarray}
where $\mu_{0}$ and $e$ denote the mass and the charge of the scalar particle, respectively. Note that we have chosen the units $G \equiv c \equiv \hbar \equiv 1$.

Thus, substituting Eqs.~(\ref{eq:metrica_KS}) and (\ref{eq:potencial_EM_KS}) into Eq.~(\ref{eq:Klein-Gordon_gauge}), we obtain
\begin{eqnarray}
&& \biggl\{-\frac{1}{\Delta_{\mbox{\tiny{KS}}}}[(r^{2}+2br+a^{2})^{2}-\Delta_{\mbox{\tiny{KS}}}a^{2}\sin^{2}\theta]\frac{\partial^{2}}{\partial t^{2}}+\frac{\partial}{\partial r}\biggl(\Delta_{\mbox{\tiny{KS}}}\frac{\partial}{\partial r}\biggr)\nonumber\\
&& +\frac{1}{\sin\theta}\frac{\partial}{\partial \theta}\biggl(\sin\theta\frac{\partial}{\partial \theta}\biggr)+\frac{\Delta_{\mbox{\tiny{KS}}}-a^{2}\sin^{2}\theta}{\Delta_{\mbox{\tiny{KS}}}\sin^{2}\theta}\frac{\partial^{2}}{\partial\phi^{2}}-\frac{4Mar}{\Delta_{\mbox{\tiny{KS}}}}\frac{\partial^{2}}{\partial t\partial\phi}\nonumber\\
&& -\frac{2ieQr}{\Delta_{\mbox{\tiny{KS}}}}\biggl[(r^{2}+2br+a^{2})\frac{\partial}{\partial t}+a\frac{\partial}{\partial\phi}\biggr]+\frac{e^{2}Q^{2}r^{2}}{\Delta_{\mbox{\tiny{KS}}}}-\mu_{0}^{2}\rho_{\mbox{\tiny{KS}}}^{2}\biggr\}\Psi=0\ .
\label{eq:mov_1_KS}
\end{eqnarray}

At this point, we need to choose a separation of variables for the scalar wave function $\Psi$ that permits us to separate the radial and angular parts of the Klein-Gordon equation (\ref{eq:mov_1_KS}). In this manner, by taking into account the spacetime symmetry, we can write $\Psi$ as
\begin{equation}
\Psi=\Psi(\mathbf{r},t)=R(r)S(\theta)\mbox{e}^{im\phi}\mbox{e}^{-i \omega t}\ ,
\label{eq:separacao_variaveis_KS}
\end{equation}
where $m$ is the azimuthal quantum number and $\omega$ is the frequency (energy) of the scalar particle.

In what follows, we will solve both the angular and radial equations of the covariant Klein-Gordon equation.
%
%
\subsection{Angular equation}
Thus, by substituting of Eq.~(\ref{eq:separacao_variaveis_KS}) into (\ref{eq:mov_1_KS}), we obtain the angular part of the covariant Klein-Gordon equation, namely,
\begin{equation}
\frac{1}{\sin\theta}\frac{d}{d\theta}\biggl(\sin\theta\frac{dS}{d\theta}\biggr)+\biggl(\Lambda_{m}+c_{0}^{2}\cos^{2}\theta-\frac{m^{2}}{\sin^{2}\theta}\biggr)S=0\ ,
\label{eq:mov_angular_1_KS}
\end{equation}
with
\begin{equation}
c_{0}^{2}=a^{2}(\omega^{2}-\mu_{0}^{2})\ ,
\label{eq:c_mov_angular_1_KS}
\end{equation}
\begin{equation}
\Lambda_{m}=\lambda_{m}-c_{0}^{2}+2a\omega m\ ,
\label{eq:Lambda_mov_angular_1_KS}
\end{equation}
where $\lambda_{m}$ is a separation constant.

Thus, to solve Eq.~(\ref{eq:mov_angular_1_KS}) we use the procedure developed in our recent papers (see for example \cite{JCAP01(2018)006} and references therein). In this manner, the general solution of the angular part of the covariant Klein-Gordon equation for a charged massive scalar field in the KS black hole, in the region exterior to the event horizon, can be written as
\begin{eqnarray}
S(x) & = & x^{\frac{1}{2}\left(\frac{1}{2}+\beta\right)}(x-1)^{\frac{1}{2}\gamma}\{C_{1}\ \mbox{HeunC}(\alpha,\beta,\gamma,\delta,\eta;x)\nonumber\\
&& +C_{2}\ x^{-\beta}\ \mbox{HeunC}(\alpha,-\beta,\gamma,\delta,\eta;x)\}\ ,
\label{eq:solucao_geral_angular_KS}
\end{eqnarray}
with
\begin{equation}
x=\cos^{2}\theta\ ,
\label{eq:x_angular_KS}
\end{equation}
where $C_{1}$ and $C_{2}$ are constants, and the parameters $\alpha$, $\beta$, $\gamma$, $\delta$, and $\eta$ are given by
\begin{equation}
\alpha=0\ ,
\label{eq:alpha_angular_HeunC_KS}
\end{equation}
\begin{equation}
\beta=\frac{1}{2}\ ,
\label{eq:beta_angular_HeunC_KS}
\end{equation}
\begin{equation}
\gamma=m\ ,
\label{eq:gamma_angular_HeunC_KS}
\end{equation}
\begin{equation}
\delta=-\frac{c_{0}^{2}}{4}\ ,
\label{eq:delta_angular_HeunC_KS}
\end{equation}
\begin{equation}
\eta=\frac{1}{4}(1+m^{2}-\Lambda_{m})\ .
\label{eq:eta_angular_HeunC_KS}
\end{equation}
Due to the fact that $\beta$ is not necessarily an integer, these two functions form linearly independent solutions of the confluent Heun dif\-fer\-en\-tial equation \cite{JPhysAMathTheor.43.035203}, namely,
\begin{equation}
\frac{d^{2}U(x)}{dx^{2}}+\biggl(\alpha+\frac{\beta+1}{x}+\frac{\gamma+1}{x-1}\biggr)\frac{dU(x)}{dx}+\biggl(\frac{\mu}{x}+\frac{\nu}{x-1}\biggr)U(x)=0\ ,
\label{eq:Heun_confluente_forma_canonica}
\end{equation}
where $U(x)=\mbox{HeunC}(\alpha,\beta,\gamma,\delta,\eta;x)$ is the confluent Heun function. The parameters $\mu$ and $\nu$ are given by
\begin{equation}
\mu=\frac{1}{2}(\alpha-\beta-\gamma+\alpha\beta-\beta\gamma)-\eta\ ,
\label{eq:mu_Heun_conlfuente_2}
\end{equation}
\begin{equation}
\nu=\frac{1}{2}(\alpha+\beta+\gamma+\alpha\gamma+\beta\gamma)+\delta+\eta\ .
\label{eq:nu_Heun_conlfuente_2}
\end{equation}
%
%
\subsection{Radial equation}
The radial part of the covariant Klein-Gordon equation can be written as
\begin{eqnarray}
&& \frac{d}{dr}\biggl(\Delta_{\mbox{\tiny{KS}}}\frac{dR}{dr}\biggr)+\biggl\{\frac{1}{\Delta_{\mbox{\tiny{KS}}}}[\omega(r^{2}+2br+a^{2})-(am+eQr)]^{2}\nonumber\\
&& -[\lambda_{m}+\mu_{0}^{2}(r^{2}+2br+a^{2})]\biggr\}R=0\ .
\label{eq:mov_radial_1_KS}
\end{eqnarray}

The general solution of Eq.~(\ref{eq:mov_radial_1_KS}), in the region exterior to the event horizon, can be written as
\begin{eqnarray}
R(x) & = & x^{\frac{\beta}{2}}(x-1)^{\frac{\gamma}{2}}\mbox{e}^{\frac{\alpha}{2}x}\{C_{1}\ \mbox{HeunC}(\alpha,\beta,\gamma,\delta,\eta;x)\nonumber\\
&& +C_{2}\ x^{-\beta}\ \mbox{HeunC}(\alpha,-\beta,\gamma,\delta,\eta;x)\}\ ,
\label{eq:solucao_geral_radial_KS}
\end{eqnarray}
where we have introduced a new variable defined by
\begin{equation}
x=\frac{r-r_{+}}{r_{-}-r_{+}}\ .
\label{eq:x_KS}
\end{equation}
In Eq.~(\ref{eq:solucao_geral_radial_KS}), $C_{1}$ and $C_{2}$ are constants, and the parameters $\alpha$, $\beta$, $\gamma$, $\delta$, and $\eta$ are now given by
\begin{equation}
\alpha=2i(r_{+}-r_{-})(\omega^{2}-\mu_{0}^{2})^{\frac{1}{2}}\ ,
\label{eq:alpha_radial_HeunC_KS}
\end{equation}
\begin{equation}
\beta=\frac{2i}{r_{+}-r_{-}}[\omega(r^{2}_{+}+2br_{+}+a^{2})-(am+eQr_{+})]\ ,
\label{eq:beta_radial_HeunC_KS}
\end{equation}
\begin{equation}
\gamma=\frac{2i}{r_{+}-r_{-}}[\omega(r^{2}_{-}+2br_{-}+a^{2})-(am+eQr_{-})]\ ,
\label{eq:gamma_radial_HeunC_KS}
\end{equation}
\begin{equation}
\delta=(r_{+}-r_{-})[2eQ\omega+(r_{+}+r_{-}+2b)(\mu_{0}^{2}-2\omega^{2})]\ ,
\label{eq:delta_radial_HeunC_KS}
\end{equation}
\begin{eqnarray}
\eta & = & -\frac{1}{(r_{+}-r_{-})^2}\{2 e^2 Q^2 r_{+} r_{-}+r_{+}^2 \lambda_{m} -2 r_{+} r_{-} \lambda_{m} +r_{-}^2 \lambda_{m} +2 b r_{+}^3 \mu_{0} ^2\nonumber\\
& & +r_{+}^4 \mu_{0} ^2-4 b r_{+}^2 r_{-} \mu_{0} ^2-2 r_{+}^3 r_{-} \mu_{0} ^2+2 b r_{+} r_{-}^2 \mu_{0} ^2+r_{+}^2 r_{-}^2 \mu_{0} ^2-4 a^3 m \omega \nonumber\\
& & +2 e Q r_{+} (r_{+}^2-4 b r_{-}-3 r_{+} r_{-}) \omega +2 a^4 \omega ^2-4 b r_{+}^3 \omega ^2-2 r_{+}^4 \omega ^2\nonumber\\
& & +8 b^2 r_{+} r_{-} \omega ^2+12 b r_{+}^2 r_{-} \omega ^2+4 r_{+}^3 r_{-} \omega ^2\nonumber\\
& & +2 a m [(e Q -2\omega b)(r_{+}+r_{-})-2\omega r_{+} r_{-}]\nonumber\\
& & +a^2 [2 m^2+(r_{+}^2+r_{-}^2) \mu_{0} ^2+r_{-} (4 b \omega ^2-2 e Q \omega )\nonumber\\
& & -2 r_{+} (r_{-} \mu_{0} ^2+e Q \omega -2 b \omega ^2-2 r_{-} \omega ^2)]\}\ .
\label{eq:eta_radial_HeunC_KS}
\end{eqnarray}

Next, we will use this radial solution to investigate two interesting physical phenomena: the black hole radiation and the discrete spectrum of energy levels.
%
%
\section{Hawking radiation}\label{Sec.III}
In this section, we will use the radial solution and the properties of the confluent Heun function in order to obtain the Hawking radiation spectrum.

The expansion in power series of the confluent Heun function with respect to the independent variable $x$, in a neighborhood of the regular singular point $x=0$, is given by \cite{Ronveaux:1995}
\begin{eqnarray}
\mbox{HeunC}(\alpha,\beta,\gamma,\delta,\eta;x) & = & 1+\frac{1}{2}\frac{(-\alpha\beta+\beta\gamma+2\eta-\alpha+\beta+\gamma)}{(\beta+1)}x\nonumber\\
&& +\frac{1}{8}\frac{1}{(\beta+1)(\beta+2)}(\alpha^{2}\beta^{2}-2\alpha\beta^{2}\gamma+\beta^{2}\gamma^{2}\nonumber\\
&& -4\eta\alpha\beta+4\eta\beta\gamma+4\alpha^{2}\beta-2\alpha\beta^{2}-6\alpha\beta\gamma\nonumber\\
&& +4\beta^{2}\gamma+4\beta\gamma^{2}+4\eta^{2}-8\eta\alpha+8\eta\beta+8\eta\gamma\nonumber\\
&& +3\alpha^{2}-4\alpha\beta-4\alpha\gamma+3\beta^{2}+4\beta\delta\nonumber\\
&& +10\beta\gamma+3\gamma^{2}+8\eta+4\beta+4\delta+4\gamma)x^2\nonumber\\
&& +\ldots\ .
\label{eq:serie_HeunC_todo_x}
\end{eqnarray}

Thus, the radial solution when $r \rightarrow r_{+}$, which implies that $x \rightarrow 0$, behaves as
\begin{equation}
R(r) \sim C_{1}\ (r-r_{+})^{\beta/2}+C_{2}\ (r-r_{+})^{-\beta/2}\ ,
\label{eq:exp_0_solucao_geral_radial_KS}
\end{equation}
where only the contributions of the first term in the expansion were taken into account, and all remaining constants were included in $C_{1}$ and $C_{2}$. Then, by taking into account the solution of time dependence, near the KS black hole event horizon $r_{+}$, we can write
\begin{equation}
\Psi=\mbox{e}^{-i \omega t}(r-r_{+})^{\pm\beta/2}\ .
\label{eq:sol_onda_radial_KS}
\end{equation}
In this case, from Eq.~(\ref{eq:beta_radial_HeunC_KS}), for the parameter $\beta$, we obtain
\begin{equation}
\frac{\beta}{2}=\frac{i}{2\kappa_{+}}(\omega-\omega_{+})\ ,
\label{eq:beta/2_solucao_geral_radial_KS}
\end{equation}
where 
\begin{equation}
\omega_{\pm}=m\Omega_{\pm}+e\Phi_{\pm}\ .
\label{eq:omega0_KS}
\end{equation}
Thus, the ingoing and outgoing wave solutions in the KS black hole horizon surface can be written as
\begin{equation}
\Psi_{in}=\mbox{e}^{-i \omega t}(r-r_{+})^{-\frac{i}{2\kappa_{+}}(\omega-\omega_{+})}\ ,
\label{eq:sol_in_1_KS}
\end{equation}
\begin{equation}
\Psi_{out}(r>r_{+})=\mbox{e}^{-i \omega t}(r-r_{+})^{\frac{i}{2\kappa_{+}}(\omega-\omega_{+})}\ .
\label{eq:sol_out_2_KS}
\end{equation}

The relative scattering probability of the scalar wave at the exterior event horizon surface is given by
\begin{equation}
\Gamma_{+}(\omega)=\left|\frac{\Psi_{out}(r>r_{+})}{\Psi_{out}(r<r_{+})}\right|^{2}=\mbox{e}^{-\frac{2\pi}{\kappa_{+}}(\omega-\omega_{+})}=\mbox{e}^{-\beta_{+}(\omega-\omega_{+})}\ .
\label{eq:taxa_refl_KS}
\end{equation}
Thus, according to the Damour-Ruffini-Sannan method \cite{PhysRevD.14.332,GenRelativGravit.20.239}, we can obtain a Hawking radiation spectrum of scalar particles, which is given by
\begin{equation}
\bar{N}_{\omega}=\frac{\Gamma_{+}}{1-\Gamma_{+}}=\frac{1}{\mbox{e}^{\beta_{+}(\omega-\omega_{+})}-1}\ .
\label{eq:espectro_rad_KS_2}
\end{equation}
Equation (\ref{eq:espectro_rad_KS_2}) corresponds to the black body spectrum described by charged massive scalar particles that are emitted from the KS black hole. In fact, this is a finite solution for the wave function near the exterior event horizon of the background spacetime under consideration.

Now, we consider the limit where $\omega$ is a continuous variable. Thus, we obtain the following expression for the total rate of particle emission
\begin{equation}
\frac{dN}{dt}=\int_{0}^{\infty}\frac{\bar{N}_{\omega}}{2\pi}\ d\omega=\frac{1}{\beta_{+}}\ln\biggl(\frac{1}{1-\mbox{e}^{\beta_{+}\omega_{+}}}\biggr)\ .
\label{eq:total_rate_KS}
\end{equation}
Furthermore, the mass loss rate is given by
\begin{equation}
\frac{dM}{dt}=-\frac{1}{2\pi}\int_{0}^{\infty}\bar{N}_{\omega}\ \omega\ d\omega=-\frac{1}{2\pi}\frac{1}{\beta_{+}^{2}}\mbox{Li}_{2}(\mbox{e}^{\beta_{+}\omega_{+}})\ ,
\label{eq:Mass_loss_KS}
\end{equation}
where $\mbox{Li}_{n}(z)$ is the polylogarithm function with $n$ running from 1 to $\infty$. Therefore, the flux of charged scalar particles at infinity, $\Phi$, can be calculated as
\begin{equation}
\Phi=\left|\frac{dM}{dt}\right|=\frac{1}{2\pi}\frac{1}{\beta_{+}^{2}}\mbox{Li}_{2}(\mbox{e}^{\beta_{+}\omega_{+}})\ .
\label{eq:Flux_KS}
\end{equation}

According to the canonical assembly theory \cite{CommunTheorPhys.52.189}, assuming that the frequency (energy) is continuous, the free energy of the scalar particle is given by
\begin{equation}
F_{+}=-\int_{0}^{\infty}\frac{\Gamma_{+}(\omega)}{\mbox{e}^{\beta_{+}(\omega-\omega_{+})}-1}\ d\omega=\frac{\mbox{e}^{\beta_{+}\omega_{+}}+\ln(1-\mbox{e}^{\beta_{+}\omega_{+}})}{\beta_{+}}\ .
\label{eq:free_energy_KS}
\end{equation}

Finally, the entropy of the KS black hole is given by
\begin{equation}
S=\beta_{+}^{2}\frac{\partial F}{\partial \beta_{+}}=\frac{\mbox{e}^{\beta_{+}\omega_{+}}[1+\mbox{e}^{\beta_{+}\omega_{+}}(\beta_{+}\omega_{+}-1)]}{\mbox{e}^{\beta_{+}\omega_{+}}-1}-\ln(1-\mbox{e}^{\beta_{+}\omega_{+}})\ .
\label{eq:Entropy_system_KS}
\end{equation}

The above results show that the radiation spectrum, flux of particles, as well as the thermodynamics variables such as free energy and entropy are similar to the corresponding ones in the framework of Kerr-Sen black hole spacetime.
%
%
\section{Resonant frequencies}\label{Sec.IV}
In this section, we want to compute the field energies, that is, the resonant frequencies for scalar waves propagating in a Kerr-Sen black hole. To do this, we will follow our technique developed in Ref.~\cite{AnnPhys.373.28}.

This approach imposes boundary conditions on the radial solution, which are as follows: the radial solution should be finite on the exterior event horizon and well behaved at an asymptotic infinity.

The first condition is completely satisfied as we can see from the wave solution that describes the Hawking radiation in the background under consideration. The latter condition requires that $R(x)$ must have a polynomial form. Then, the function $\mbox{HeunC}(\alpha,\beta,\gamma,\delta,\eta;x)$ becomes a polynomial of degree $n$ if the following $\delta$-condition is satisfied
\begin{equation}
\frac{\delta}{\alpha}+\frac{\beta+\gamma}{2}+1=-n\ ,
\label{eq:delta_condition}
\end{equation}
where $n=\{0,1,2,\ldots\}$ is the principal quantum number.

Substituting Eqs.~(\ref{eq:alpha_radial_HeunC_KS})-(\ref{eq:delta_radial_HeunC_KS}) into Eq.~(\ref{eq:delta_condition}), we obtain the following expression that involves the resonant frequencies associated with charged massive scalar particles in a Kerr-Sen black holes:
\begin{eqnarray}
&& \frac{\omega e Q+M(\mu_{0}^{2}-2\omega^{2})}{(\mu_{0}^{2}-\omega^{2})^{\frac{1}{2}}}\nonumber\\
&& +\frac{i}{2}\left[\omega\left(\frac{1}{\kappa_{+}}+\frac{1}{\kappa_{-}}\right)-\left(\frac{\omega_{+}}{\kappa_{+}}+\frac{\omega_{-}}{\kappa_{-}}\right)\right]=-(n+1)\ .
\label{eq:quasispectrum_modes_KS_case2_x}
\end{eqnarray}
This quantization law gives a complex number, that is, we obtain a frequency (energy) spectrum such that $\omega=\omega_{R}+i\ \omega_{I}$, where $\omega_{R}$ and $\omega_{I}$ are the real and imaginary parts, respectively. Note that there is no dependence on the eigenvalue $\lambda_{m}$ and hence the eigenvalues given by Eq.~(\ref{eq:quasispectrum_modes_KS_case2_x}) are not degenerate.

Equation (\ref{eq:quasispectrum_modes_KS_case2_x}) is a second-order equation for $\omega$ and hence has two solutions, which can be numerically obtained using the \textit{FindRoot} function in the \textbf{Wolfram Mathematica}$^{\mbox{\tiny\textregistered}}$ \textbf{9}, such that $(\omega-\omega^{\ (1)}_{n})(\omega-\omega^{\ (2)}_{n})=0$. The resonant frequencies $\omega^{\ (1)}_{0}$ and $\omega^{\ (2)}_{0}$ are shown in Tables \ref{tab:Rfs_KS_1} and \ref{tab:Rfs_KS_2}, respectively, where the units are given as multiples of the total mass $M$.

\begin{center}
\tabcaption{ \label{tab:Rfs_KS_1} Scalar resonant frequencies $\omega^{\ (1)}_{n}$ of a Kerr-Sen black hole for $e=0.1$, $a=0.5$, $m=0$ and $\mu_{0}=0.8$. We focus on the fundamental mode $n=0$.}
\normalsize
\begin{tabular*}{170mm}{@{\extracolsep{\fill}}lllll@{\extracolsep{\fill}}}
\toprule
			$b$  & $\mbox{Re}[\omega^{\ (1)}_{0}]$ & $\mbox{Im}[\omega^{\ (1)}_{0}]$	\\\hline
			0.00 & \ 0.00000  & 0.08126 \\
			0.01 & \ 0.00680  & 0.08101 \\
			0.04 & \ 0.01361  & 0.08018 \\
			0.09 & \ 0.02045  & 0.07861 \\
			0.16 & \ 0.02733  & 0.07582 \\
			0.25 & \ 0.03431  & 0.07074 \\
			0.36 & \ 0.04151  & 0.05995 \\
			0.49 & \ 0.04938  & 0.01948 \\
			0.64 & -0.04467   & 0.00000 \\
\bottomrule
\end{tabular*}
\end{center}

\begin{center}
\tabcaption{ \label{tab:Rfs_KS_2} Scalar resonant frequencies $\omega^{\ (2)}_{n}$ of a Kerr-Sen black hole for $e=0.1$, $a=0.5$, $m=0$ and $\mu_{0}=0.8$. We focus on the fundamental mode $n=0$.}
\normalsize
\begin{tabular*}{170mm}{@{\extracolsep{\fill}}lllll@{\extracolsep{\fill}}}
\toprule
			$b$  & $\mbox{Re}[\omega^{\ (2)}_{0}]$ & $\mbox{Im}[\omega^{\ (2)}_{0}]$	\\\hline
			0.00 & -0.80687 & 0.07081 \\
			0.01 & -0.80781 & 0.07236 \\
			0.04 & -0.80977 & 0.07310 \\
			0.09 & -0.81309 & 0.07252 \\
			0.16 & -0.81824 & 0.06926 \\
			0.25 & -0.82530 & 0.05996 \\
			0.36 & -0.82942 & 0.03705 \\
			0.49 & -0.80429 & 0.00103 \\
			0.64 & -0.62255 & 0.00000 \\
\bottomrule
\end{tabular*}
\end{center}

The resonant frequencies obtained are shown in Figs.~\ref{fig:Fig1_KS}, \ref{fig:Fig2_KS}, \ref{fig:Fig3_KS}, \ref{fig:Fig4_KS}, \ref{fig:Fig5_KS}, and \ref{fig:Fig6_KS} as a function of $e$, $a$, $b$, $m$, $\mu_{0}$, and $n$, respectively, where the units are given as multiples of the total mass $M$.

\begin{center}
		\includegraphics[scale=1.00]{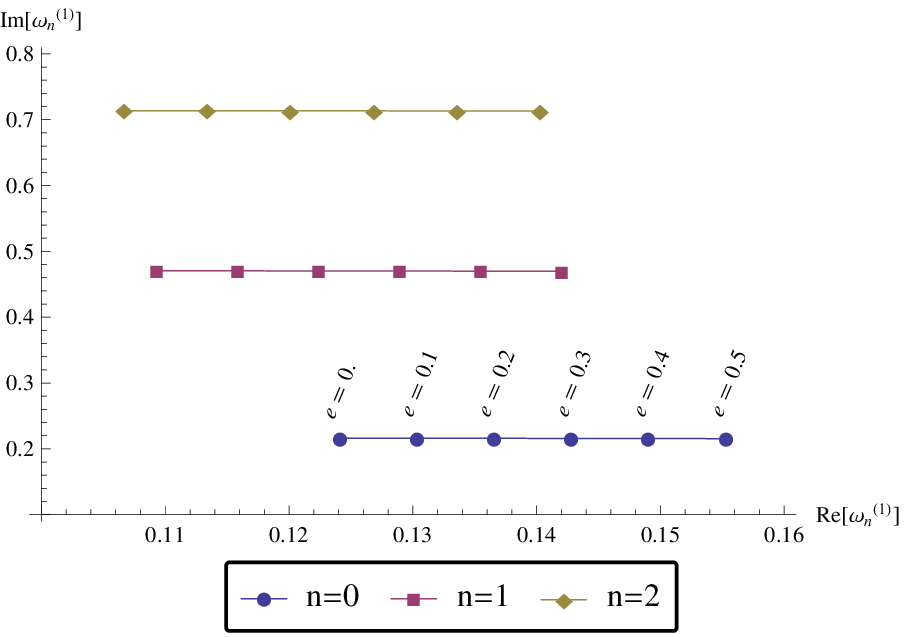}
	\figcaption{Scalar resonant frequencies of a Kerr-Sen black hole as a function of $e$ for $a=0.4$, $b=0.01$, $m=1$, and $\mu_{0}=0.4$.}
	\label{fig:Fig1_KS}
%
		\includegraphics[scale=1.00]{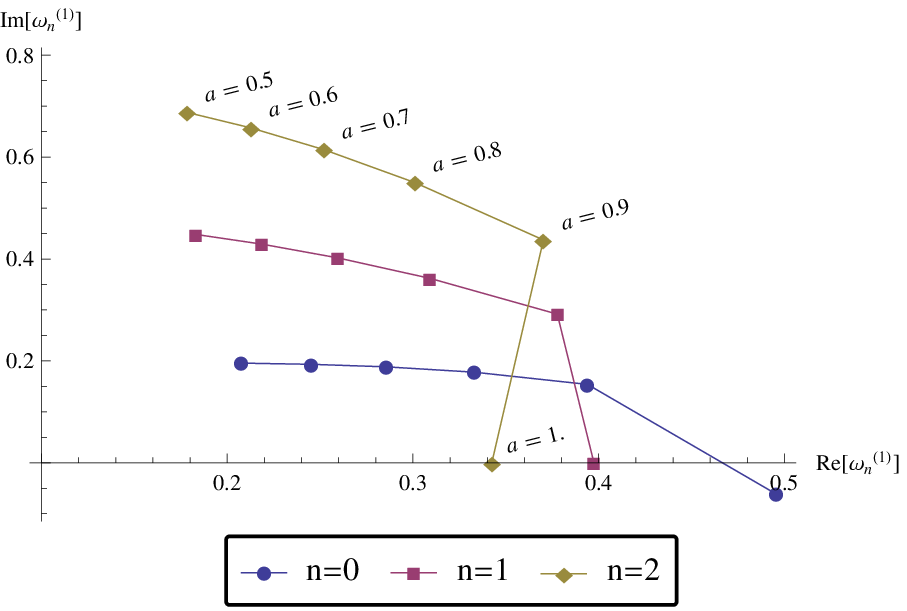}
	\figcaption{Scalar resonant frequencies of a Kerr-Sen black hole as a function of $a$ for $e=0.6$, $b=0.01$, $m=1$, and $\mu_{0}=0.5$.}
	\label{fig:Fig2_KS}
%
		\includegraphics[scale=1.00]{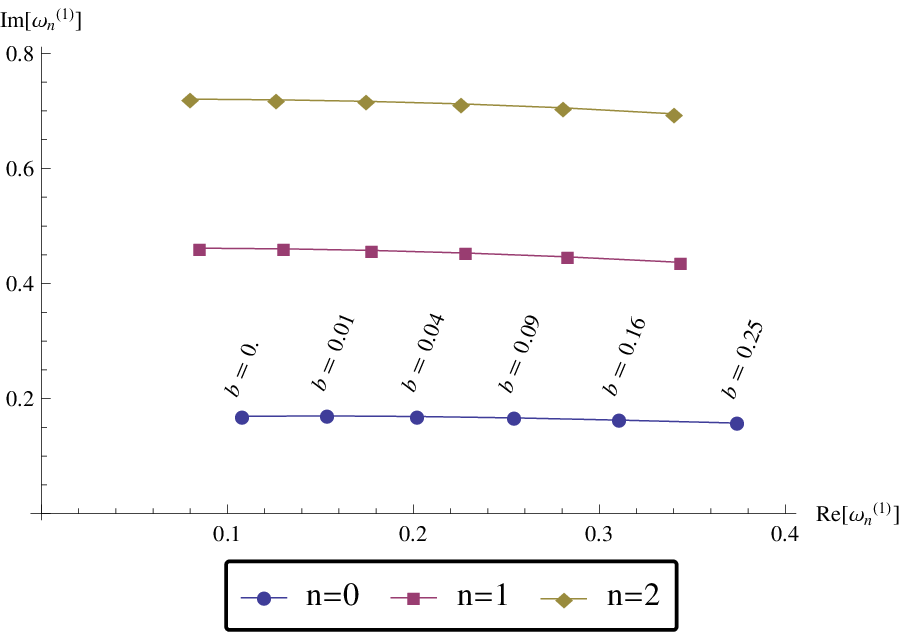}
	\figcaption{Scalar resonant frequencies of a Kerr-Sen black hole as a function of $b$ for $e=0.7$, $a=0.3$, $m=1$, and $\mu_{0}=0.6$.}
	\label{fig:Fig3_KS}
%
		\includegraphics[scale=1.00]{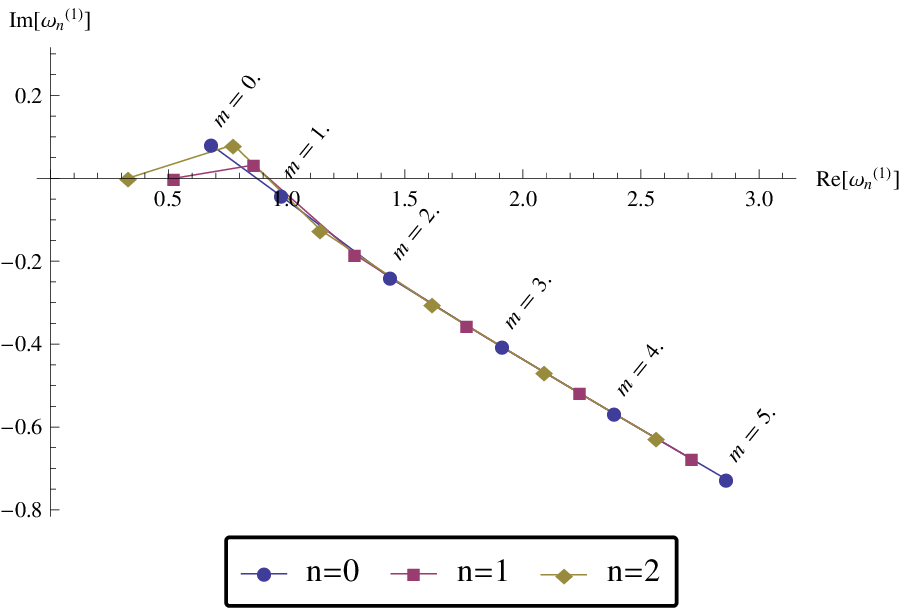}
	\figcaption{Scalar resonant frequencies of a Kerr-Sen black hole as a function of $m$ for $e=1.0$, $a=0.2$, $b=1$, and $\mu_{0}=0.7$.}
	\label{fig:Fig4_KS}
%
		\includegraphics[scale=1.00]{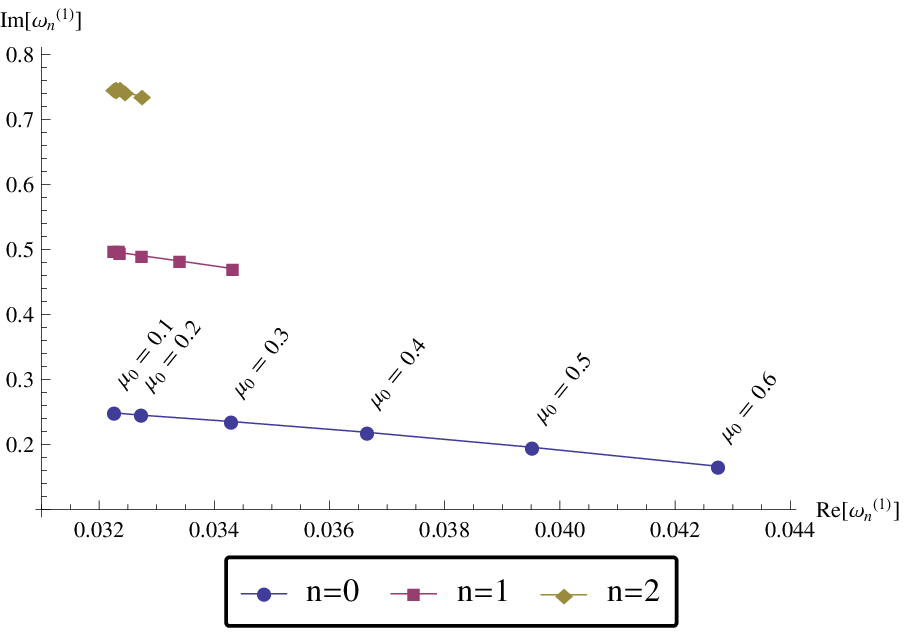}
	\figcaption{Scalar resonant frequencies of a Kerr-Sen black hole as a function of $\mu_{0}$ for $e=0.1$, $a=0.1$, $b=0.01$, and $m=1$.}
	\label{fig:Fig5_KS}
%
		\includegraphics[scale=1.00]{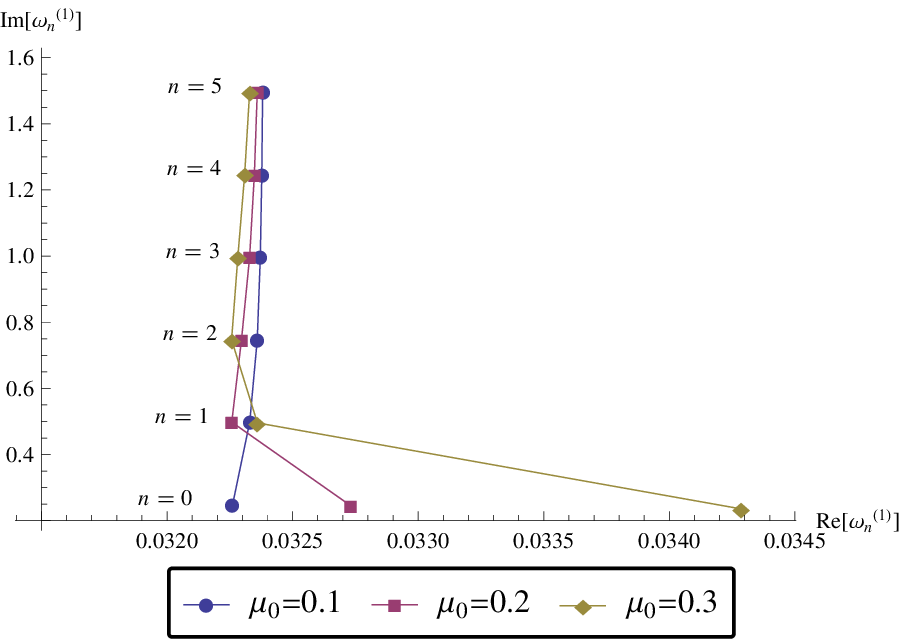}
	\figcaption{Scalar resonant frequencies of a Kerr-Sen black hole as a function of $n$ for $e=0.1$, $a=0.1$, $b=0.01$, and $m=1$.}
	\label{fig:Fig6_KS}
\end{center}
%
%
\subsection{Massless scalar fields}
The expression for the resonant frequencies can be exactly solved for $\omega_{n}$ when we have a massless scalar field. Thus, in the case where $\mu_{0}=0$, the resonant frequencies are given by
\begin{eqnarray}
\omega_{n} & = & \frac{a m-e\{\sqrt{2b^3} - M\sqrt{2b} - \sqrt{2b[(b-M)^2-a^2]}\}}{2 M (\sqrt{(b-M)^2-a^2}-b+M)}\nonumber\\
&& +i\frac{(n+1) \sqrt{(b-M)^2-a^2}}{2 M (\sqrt{(b-M)^2-a^2}-b+M)}\ ,
\label{eq:massless_Rfs_KS}
\end{eqnarray}
where the principal quantum number $n$ is either a positive integer or zero.

The eigenvalues given by Eq.~(\ref{eq:massless_Rfs_KS}) are also not degenerate as that there is no dependence on the eigenvalue $\lambda_{m}$. The massless scalar resonant frequencies, $\omega_{1}$, are shown in Table \ref{tab:Rfs_KS_3}, where the units are given as multiples of the total mass $M$.

\begin{center}
\tabcaption{ \label{tab:Rfs_KS_3} Massless scalar resonant frequencies for a Kerr-Sen black hole for $e=0.1$, $a=0.5$, and $m=0$. We focus on the first excited mode $n=1$.}
\normalsize
\begin{tabular*}{170mm}{@{\extracolsep{\fill}}lllll@{\extracolsep{\fill}}}
\toprule
			$b$  & $\mbox{Re}(\omega_{1})$ & $\mbox{Im}(\omega_{1})$	\\\hline
			0.00 & \ 0.00000 & 0.46410 \\
			0.01 & \ 0.00707 & 0.46326 \\
			0.04 & \ 0.01414 & 0.46053 \\
			0.09 & \ 0.02121 & 0.45520 \\
			0.16 & \ 0.02828 & 0.44554 \\
			0.25 & \ 0.03536 & 0.42705 \\
			0.36 & \ 0.04243 & 0.38432 \\
			0.49 & \ 0.04950 & 0.16462 \\
			0.64 & -0.44309  & 0.48160 \\
\bottomrule
\end{tabular*}
\end{center}

We present the massless scalar resonant frequencies in Figs.~\ref{fig:Fig7_KS}, \ref{fig:Fig8_KS}, \ref{fig:Fig9_KS}, \ref{fig:Fig10_KS}, and \ref{fig:Fig11_KS} as a function of $e$, $a$, $b$, $m$, and $n$, respectively, where the units are given as multiples of the total mass $M$.

\begin{center}
		\includegraphics[scale=1.00]{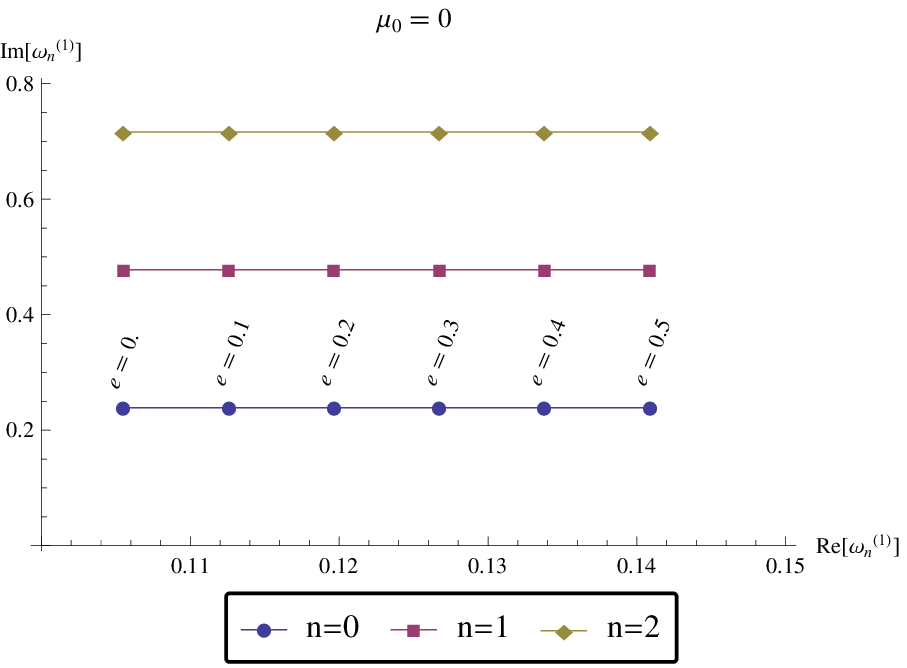}
	\figcaption{The massless scalar resonant frequencies of a Kerr-Sen black hole as a function of $e$ for $a=0.4$, $b=0.01$ and $m=1$.}
	\label{fig:Fig7_KS}
%
		\includegraphics[scale=1.00]{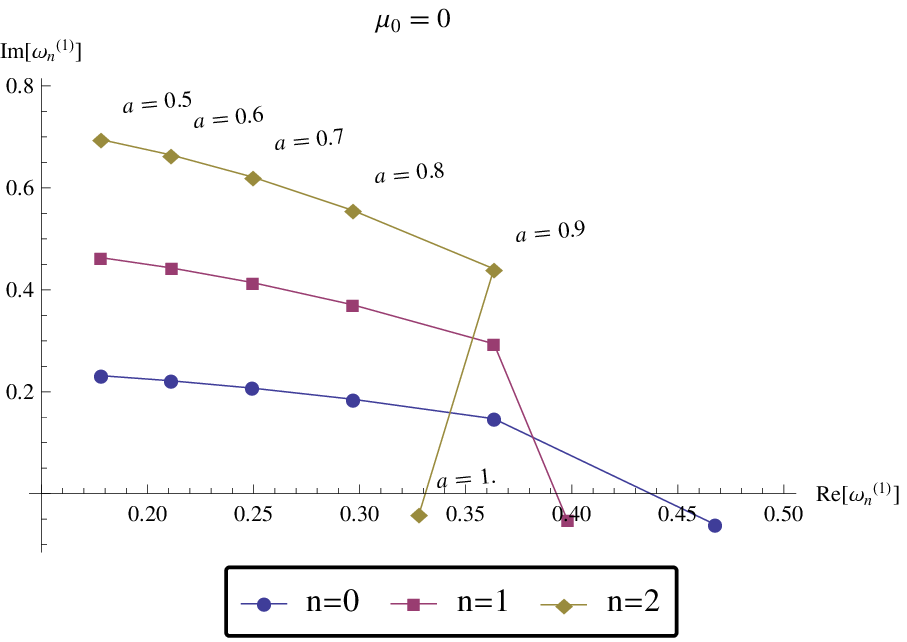}
	\figcaption{The massless scalar resonant frequencies of a Kerr-Sen black hole as a function of $a$ for $e=0.6$, $b=0.01$ and $m=1$.}
	\label{fig:Fig8_KS}
%
		\includegraphics[scale=1.00]{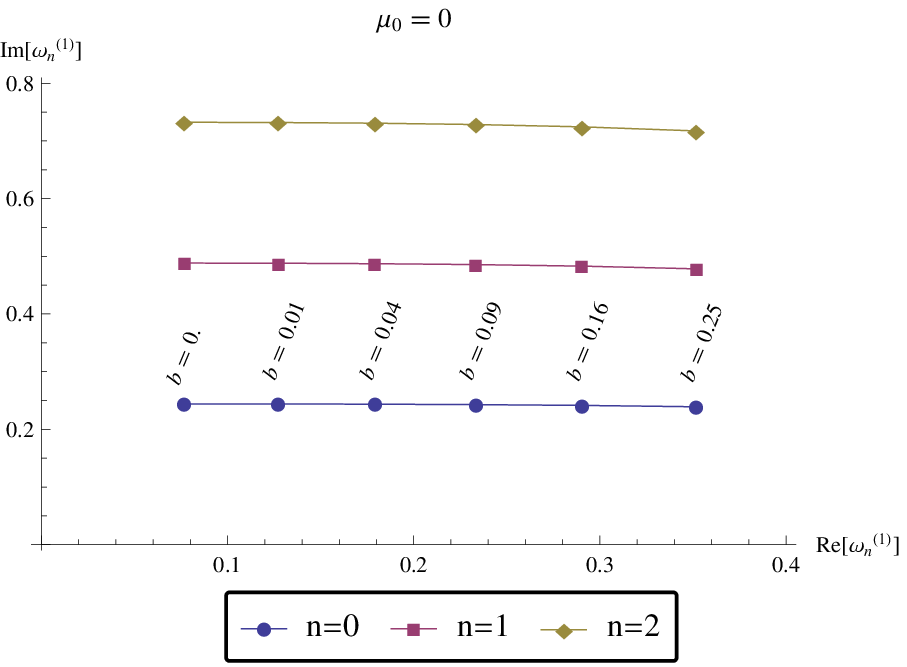}
	\figcaption{The massless scalar resonant frequencies of a Kerr-Sen black hole as a function of $b$ for $e=0.7$, $a=0.3$ and $m=1$.}
	\label{fig:Fig9_KS}
%
		\includegraphics[scale=1.00]{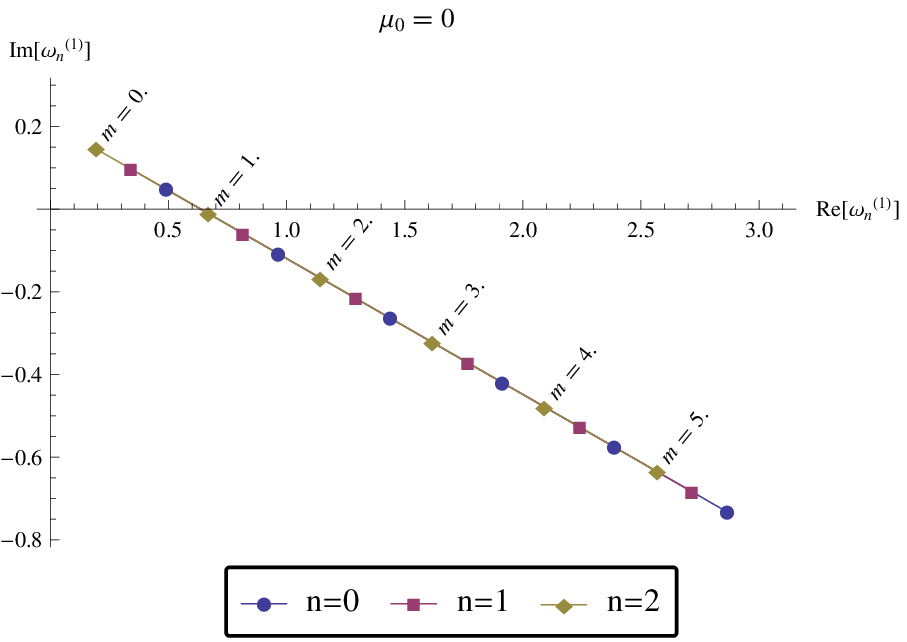}
	\figcaption{The massless scalar resonant frequencies of a Kerr-Sen black hole as a function of $m$ for $e=1.0$, $a=0.2$ and $b=1$.}
	\label{fig:Fig10_KS}
%
		\includegraphics[scale=1.00]{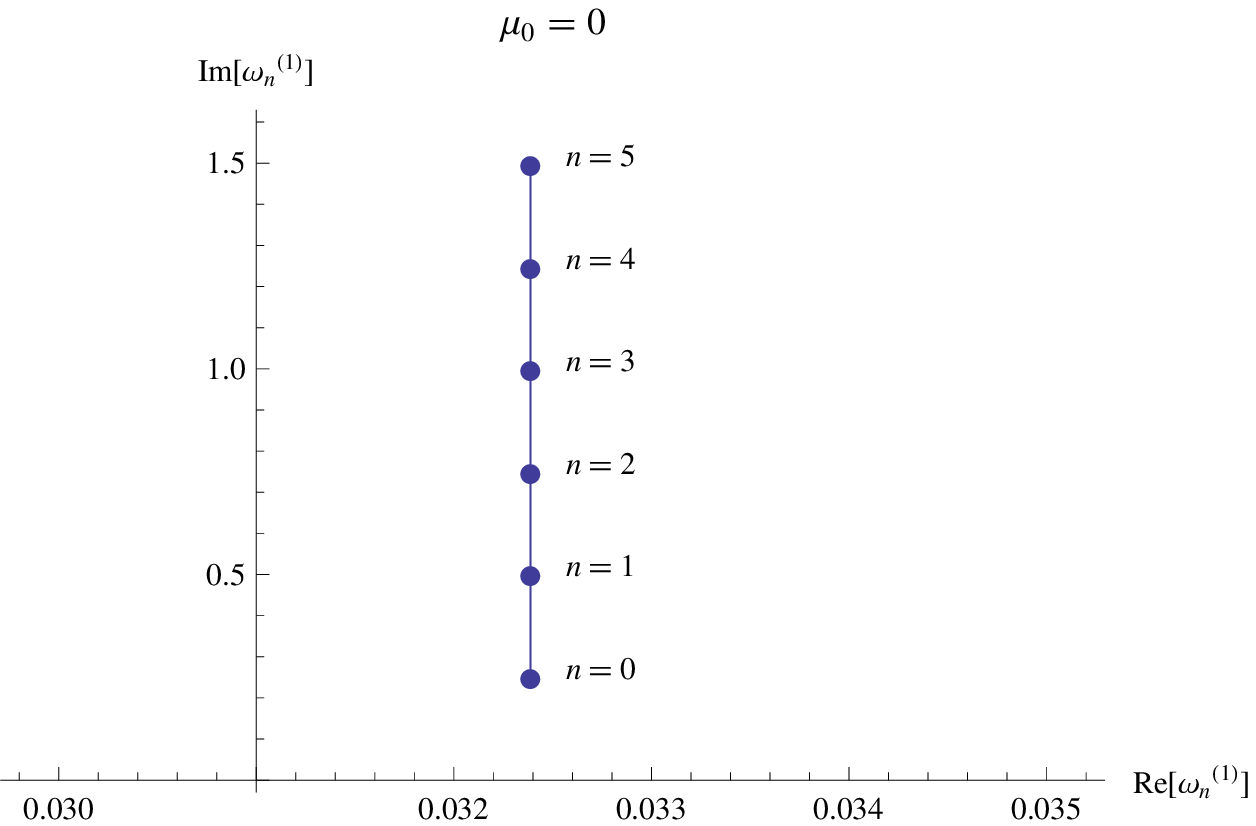}
	\figcaption{Scalar resonant frequencies of a Kerr-Sen black hole as a function of $n$ for $e=0.1$, $a=0.1$, $b=0.01$ and $m=1$.}
	\label{fig:Fig11_KS}
\end{center}
%
%
\section{Conclusions}\label{Sec.V}
In this study, we have considered the interaction between charged massive scalar fields and the Kerr-Sen black hole. We solved the covariant Klein-Gordon equation and then analyzed some interesting physical phenomena that correspond to the Hawking radiation spectrum and the resonant frequencies. We also examined how thermodynamics quantities such as entropy and free energy depend on the parameter $b$ associated to the dilaton.

The Hawking radiation spectrum was obtained from the asymptotic behavior of the radial solution at the exterior event horizon, where we have used the expansion in the power series of the confluent Heun function. With regard to the Hawking radiation we obtained a black hole radiation spectrum that resembles the one obtained in the context of the Kerr-Newman black hole spacetime. All others quantities such as the flux of particles, free energy, and entropy preserve the similarity with the corresponding results in Kerr-Newman spacetime.

We have obtained a general expression for the resonant frequencies from the boundary conditions imposed to the radial solution and studied the behavior of the oscillations and how fast they disappear. By using a numerical method, we obtain some values for the resonant frequencies as a function of the involved parameters. We also analyzed the case of massless scalar particles.

As we can see from Figs.~\ref{fig:Fig1_KS}-\ref{fig:Fig6_KS} for the massive case as well from Figs.~\ref{fig:Fig7_KS}-\ref{fig:Fig11_KS} for the massless case, the resonant frequencies depend on the parameter that codifies the presence of the dilaton field as well as on the physical parameters such as mass, charge, and angular momentum. We also emphasize how different are the resonant frequencies for different modes, for example, the fundamental and the first and second excited modes.

\vspace{15mm}

%
%


\vspace{-1mm}
\centerline{\rule{80mm}{0.1pt}}
\vspace{2mm}


%
%


\clearpage
\end{CJK*}
\end{document}